# Strong Electronic Interaction and Signatures of Nodal Superconductivity in $Zr_5Pt_3C_y$

S.T. Renosto[1], R. Lang[2], A.L.R. Manesco[3], D. Rodrigues Jr.[3], F.B. Santos[3], A.J.S. Machado[3], M.R. Baldan[4], E. Diez[1]

[1]*Universidad de Salamanca - Laboratorio de Bajas Temperaturas, Salamanca, 37008, Spain*
[2]*Universidade Federal de São Paulo - Instituto de Ciência e Tecnologia, São José dos Campos, 12231-280, Brazil*
[3]*Universidade de São Paulo – Departamento de Engenharia de Materiais, Lorena, 12602-810, Brazil*
[4]*Instituto Nacional de Pesquisas Espaciais, Laboratório Associado de Sensores e Materiais, São José dos Campos, 12227-010, Brazil*

The physical properties of the $Zr_5Pt_3$ compound with interstitial carbon in hexagonal $D8_8$ – structure was investigated. A set of macroscopic measurements reveal a bulk superconducting at approximately 7 K for $Zr_5Pt_3C_{0.3}$ close to $Zr_5Pt_3$, also with a correlate anomalous resistivity behavior. However, both the signatures of strong electron-electron interaction, and the electronic contribution to specific heat, increase dramatically with the C doping. For the first time the x-ray photoelectron spectra compared with DFT/PWLO calculations of electronic structure show a complex Fermi surface with high density of states for $Zr_5Pt_3$. Also results show the signature of unconventional superconductivity. Indeed, was observed an unusual behavior for lower and upper critical field diagrams of $Zr_5Pt_3C_{0.3}$. The temperature dependence of penetration length and electronic contribution to specific heat suggests that electronic pairing deviates of s-wave the BCS scenario.

---

In the last years, discussion about the gap-structure in unconventional superconductors (SC) come from the unusual temperature dependence of critical fields, specific heat, field penetration length, and relaxation rate of magnetic resonance [1]. In s-wave superconductors, the electron pairing is fully symmetric without low-lying collective mode since, in this case, the plasma modes respond for charged particles. The existence of isotropic gap naturally leads to an exponential temperature dependence of several SC parameters. However, the pairing symmetry may have point or line zeroes, called nodes [2], or different SC gaps on the Fermi surface (multi-band) [3]. The nodal pairing symmetry has been intensively studied by different focus; macroscopic-formalism in BCS arguments [4], and mainly with mesoscopic approach [5,6], for instance. Indeed, on these unconventional superconductors, the $T$ power-laws applied to $T << T_c$ measurements data generally distinguish the pairing symmetry: s-wave or nodal/multi-gap scenario [7]. This represents an intriguing field in contemporary physics.

In the specific case of the $D8_8$-type compounds, the superconductivity mechanism never been explored, particularly for $Zr_5Pt_3$. The interesting of $D8_8$ or $Mn_5Si_3$-prototype [space group $D^3_{6h}$–$P6_3/mcm$] is a formation of octahedral chains with internal site able to accommodate C, B, O, and N interstitial elements (X) [8-15]. In quasi-binary $Me_5M_3X_y$ compounds, the atomic positions for Me atoms are *6g* for $^3/_5$ of their atoms and *4d* for the remaining $^2/_5$; *6g* and *2b*–interstitial for M and X atoms, respectively [16-19]. Not all the *2b*–sites (blue sphere) are filled, and the Me atoms in the *4d* positions (yellow spheres) constitute the one-dimension channel [20] [Fig. 1].

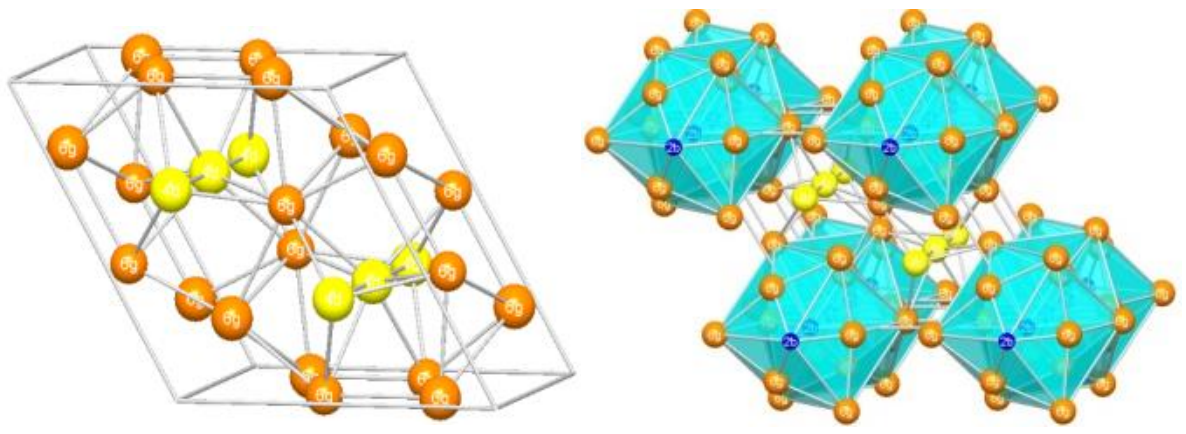

**Fig. 1:** *$D8_8$* structure with the view aligned parallel to the c-axes. On the right, it is represented the interstitial 2b–site inside the polyhedron, with triangular faces, formed by 6 metal atoms.

Extensive studies report the mechanical properties of the $D8_8$ germanides and silicides [21-23], but investigation about physical properties remain scarce. Only for $Mn_5Si_3$, the properties and the structural dependence at low temperatures were explored. Two antiferromagnetic phases accompanied by a structural distortion were observed [24,25]. In addition, an unusual magnetic structure was found to be coherent with the complex stability of Mn moments [26-28]. Nevertheless, in 2012, it was stated superconductivity in $Nb_5Ge_3$ with C interstitial doping [29]. The study raised the question about how the defect-structure influences the electronic-structure, electron-phonon interaction and structural stability. The existence of $D8_8$ compounds is well established in more than 500 systems [30]. For example, it was describe the existence of $D8_8$ in a wide variety of plumbides ($Me_5Pb_3$), stannides ($Me_5Sn_3$) and bismutides ($Me_5Bi_3$) [31-33], and in compounds based in *d quasi-complete* elements, such as $Zr_5Ir_3$ and $Zr_5Pt_3$ [34].

Specifically for $Zr_5Pt_3$ synthesized using Zr with 3% of Hf, i.e., $Zr_{4.85}Hf_{0.15}Pt_3$, superconductivity with critical temperature ($T_c$) around 7.5 K was reported in 1990 by Waterstrat *et al.* [35]. Effects by Hf contamination were evident in both, increasing $T_c$ and the lattices parameters (LP). Surprisingly, an unusual specific heat behavior below $T_c$ could be seen, and it was forgotten since then. Recently, Hamamoto *et al.* showed that $T_c$ decreases by the interstitial oxygen doping on $Zr_5Pt_3$ single-phase samples [20]. From high purity elements, the $Zr_5Pt_3$ presents significantly lower LP, and the $T_c$ = 6.4 K was confirmed by magnetization and resistivity measurements. In these works on $Zr_5Pt_3$, questions on the electronic structure and the superconducting critical parameters remain open. Motivated by those questions, we started the investigation of electric, magnetic and thermal properties; electronic structure and superconducting behavior; and as example of $Nb_5Ge_3C_x$ [29], the effects of doping with electrons by C interstitial; on $D8_8$-superconductor $Zr_5Pt_3$ and $Zr_5Pt_3C_y$ compounds.

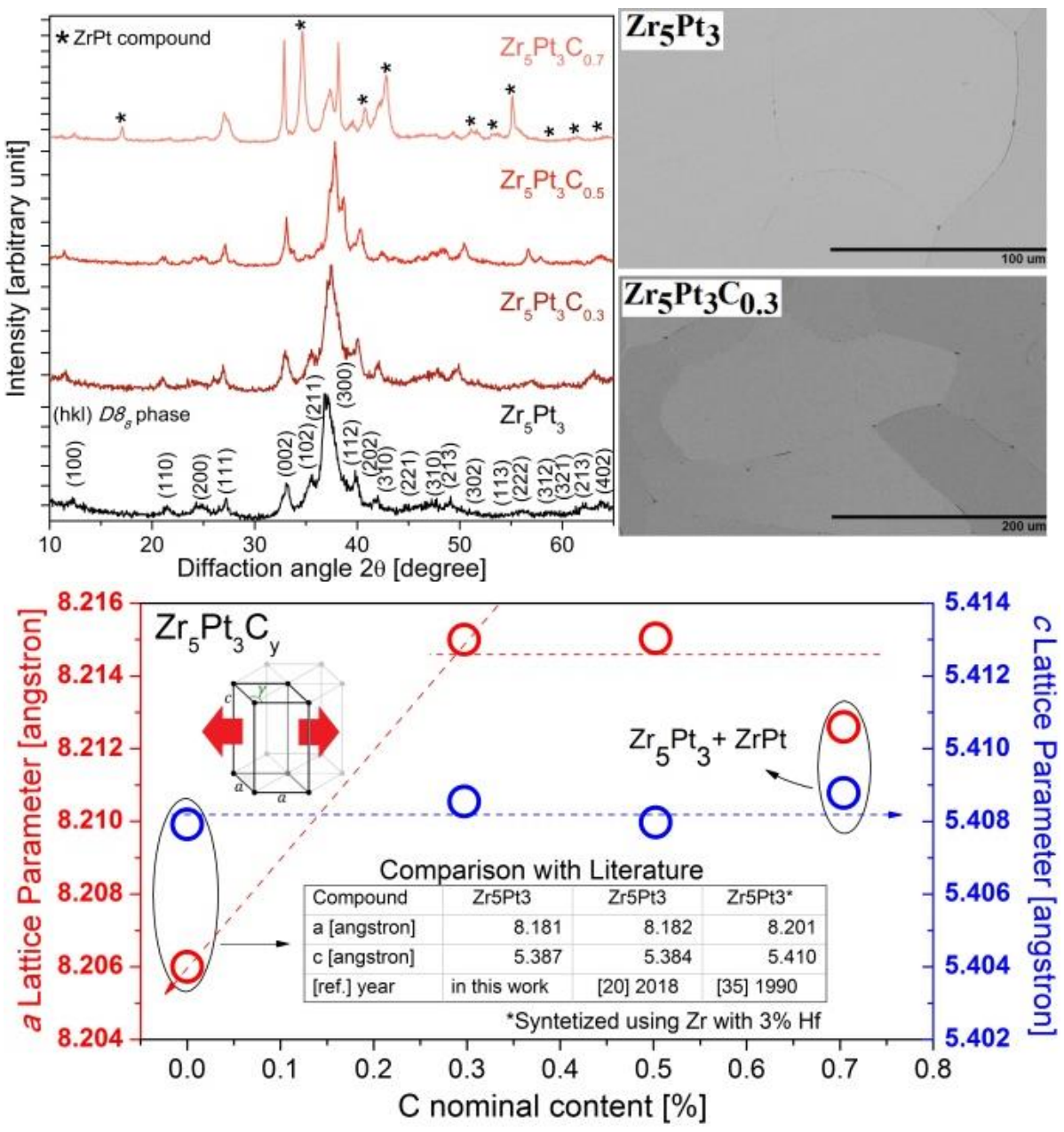

**Fig. 2:** a) XRD patterns for samples with $Zr_5Pt_3C_y$ nominal composition. (hkl) index is referent to the $D8_8$-structure. b) SEM image of $Zr_5Pt_3$ [upper panel] and $Zr_5Pt_3C_{0.3}$ [lower panel] after annealing for 180h at 1200°C. c) C-content dependence on the lattice parameters. The inset presents the comparison with the $Zr_5Pt_3$ lattice parameters reported in the literature.

$Zr_5Pt_3C_y$ ($0 \le y \le 0.7$) samples were synthesized by the arc-melting method and, due to the peritectic reaction ZrPt + liquid = $Zr_5Pt_3$, they were submitted to an annealing (1200°C/180h under Ar). Structural analysis was performed from X-ray powder diffraction (XRD) using a Panalytical Empyrean - PIXcel$^{3D}$ [Fig. 2a]. As an effect of the long annealing, it was observed a ductility increase of these compounds, which generates enlargements of the diffraction peaks after grinding. For $Zr_5Pt_3$ and $Zr_5Pt_3C_{0.3}$, the SEM/BSE micrographs, using a Hitachi-TM3000/EDS-Oxford, indicate homogeneous microstructures, with a misorientation of large grains and without the presence of secondary phases [Fig. 2b]. EDS the analyses under 20 points reveal that metallic proportion $Zr_{(4.98\pm0.01)}Pt_{(3.02\pm0.01)}$ was preserved. For C > 0.5 samples, it was observed coexistence with a secondary phase indexed as ZrPt. Consistent with C-incorporation the displacement of the (100) diffraction was observed. Indeed, results of Rietveld refinement (by GSAS software [36]) which confirm the preferable expansion of the basal plane on hexagonal unit cells [Fig. 2c]. This suggests that the interstitial doping could induce lattice dimerization along the preferable <100> and <010> direction.

Physical properties of single-phase samples were investigated by magnetization ($M$), resistivity ($\rho$), and specific heat ($C_p$) measurements in a VSM-SQUID, 9T-PPMS (Quantum Design Inc.) and He$^3$ system (Triton-Oxford). Clear diamagnetic transitions were observed in the $Zr_5Pt_3$ samples with $T_{c\text{-}onset} \approx 6$ K, and a consistent dependence on the C-content [Fig. 3a-d]. It was observed the signature of a type II superconductor with the applied magnetic field ($H$) [Fig. 3a-d inset].

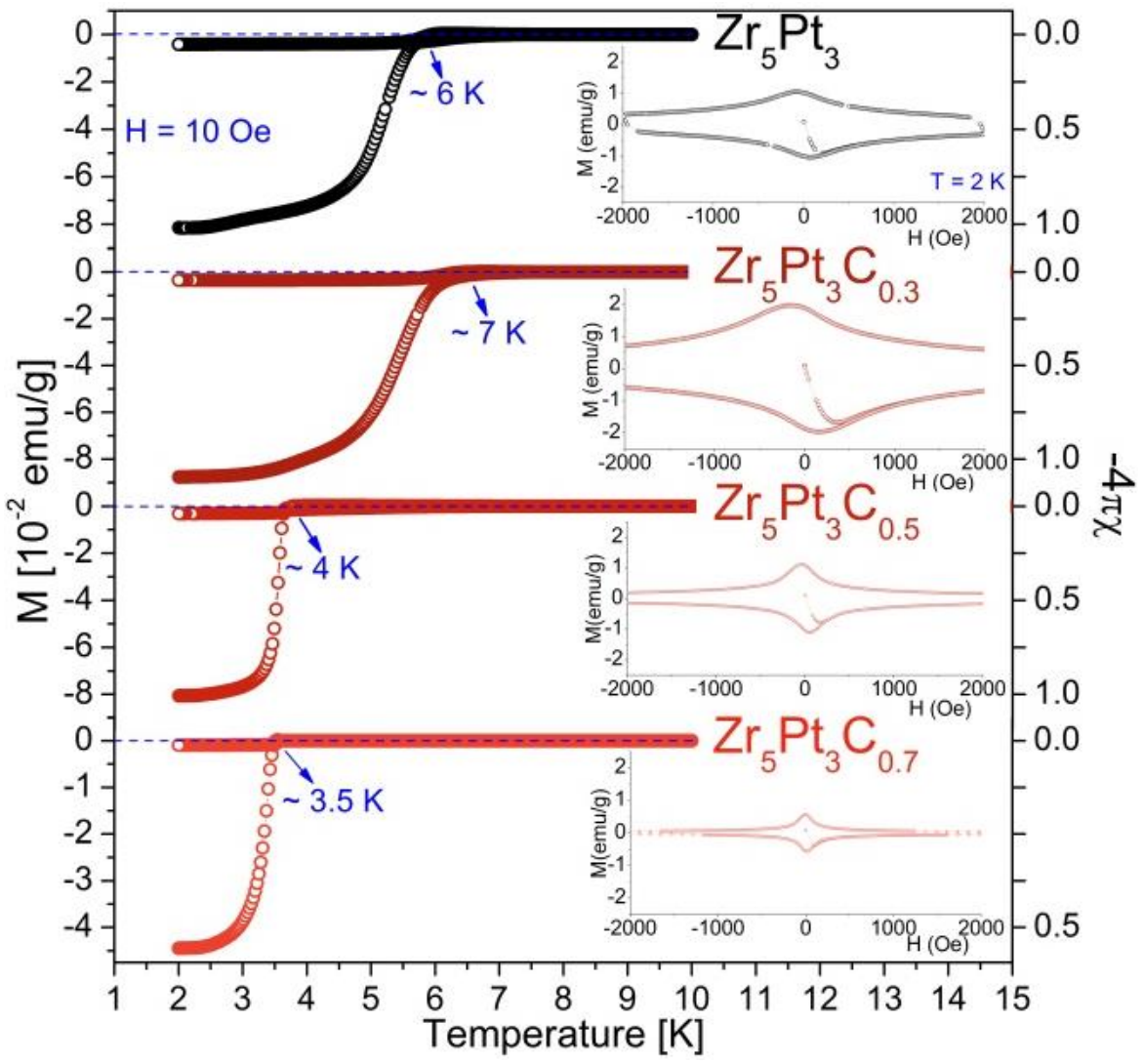

**Fig. 3a-d:** Temperature dependence of magnetization for the samples $Zr_5Pt_3$, $Zr_5Pt_3C_{0.3}$, $Zr_5Pt_3C_{0.5}$, and $Zr_5Pt_3C_{0.7}$. Insets show the dependence of the magnetization with the applied magnetic field.

A maximum value of $T_{c\text{-}onset}$, close to 7 K, was obtained for $Zr_5Pt_3C_{0.3}$, with an extreme variation of the LP [Fig. 4a]. In CGS units, the perfect diamagnetic shielding implies a magnetic susceptibility of $\chi \approx -1/4\pi$, as observed, which goes in perfect agreement with the SEM micrographs [Fig. 2b]. The linear region (Meissner line) in the $M(H)$ curve was fitted [Fig. 4a], and $\Delta M = 10^{-3}$ emu/g [37] was used as a criterion to determine the lower critical field ($H_{c1}$) [Fig. 4b]. The dependence of $H_{c1}$ with the reduced temperature ($\tau = T/T_c$) presents unusual $+T^2$ behavior (black line), that deviates from the expected for conventional superconductors [Fig. 4c]. This raises questions on the origin of superconductivity in this material. Such unusual behavior was also experimentally observed in unconventional superconductors [37-39].

A percolative superconducting path was confirmed with $\rho \rightarrow 0$ at $T_c \approx 7$ K, and with consistent displacement in the field range $0 \leq H \leq 7$ T [Fig. 5a-b]. For $Zr_5Pt_3C_{0.3}$, the Residual Resistivity Ratio (RRR) was found with the value close to 1.9, similar to RRR ≈ 2.1 obtained for $Zr_5Pt_3$ [20]. This supports the absence of electronic-scattering by a normal phase at the grain boundaries.

The mid-point of the transition was used to construct the diagram of $\tau$-dependence to the upper critical field ($\mu_0 H_{c2}$) [Fig. 5b]. Here, an unconventional slope with a positive curvature close to critical temperature was also observed [Fig. 5c]. Considering the isotropic s-wave model, $\mu_0 H_{c2}(0)$ (at zero kelvin) can be estimated using the WHH theory [$\mu_0 H_{c2}(0) = \sigma T_c \, d\mu_0 H_{c2}/dT_{(T \rightarrow Tc)}$, where $\sigma$ is 0.69 or 0.73, for the dirty and clean limit, respectively] [40]. The fitted values of $4.3 \geq \mu_0 H_{c2}(0) \geq 4.1$ T (orange line) are not suitable to describe the experimental data (black points). The opposite occurs with the fit from unusual pairing mechanism (blue line) [41-44]. In this case $\mu_0 H_{c2}(0) \approx 6.3$ T was more consistent in front of the experimental results, and conducted to a coherence length at zero kelvin ($\xi_{(T=0)}$) close to 7.1 nm.

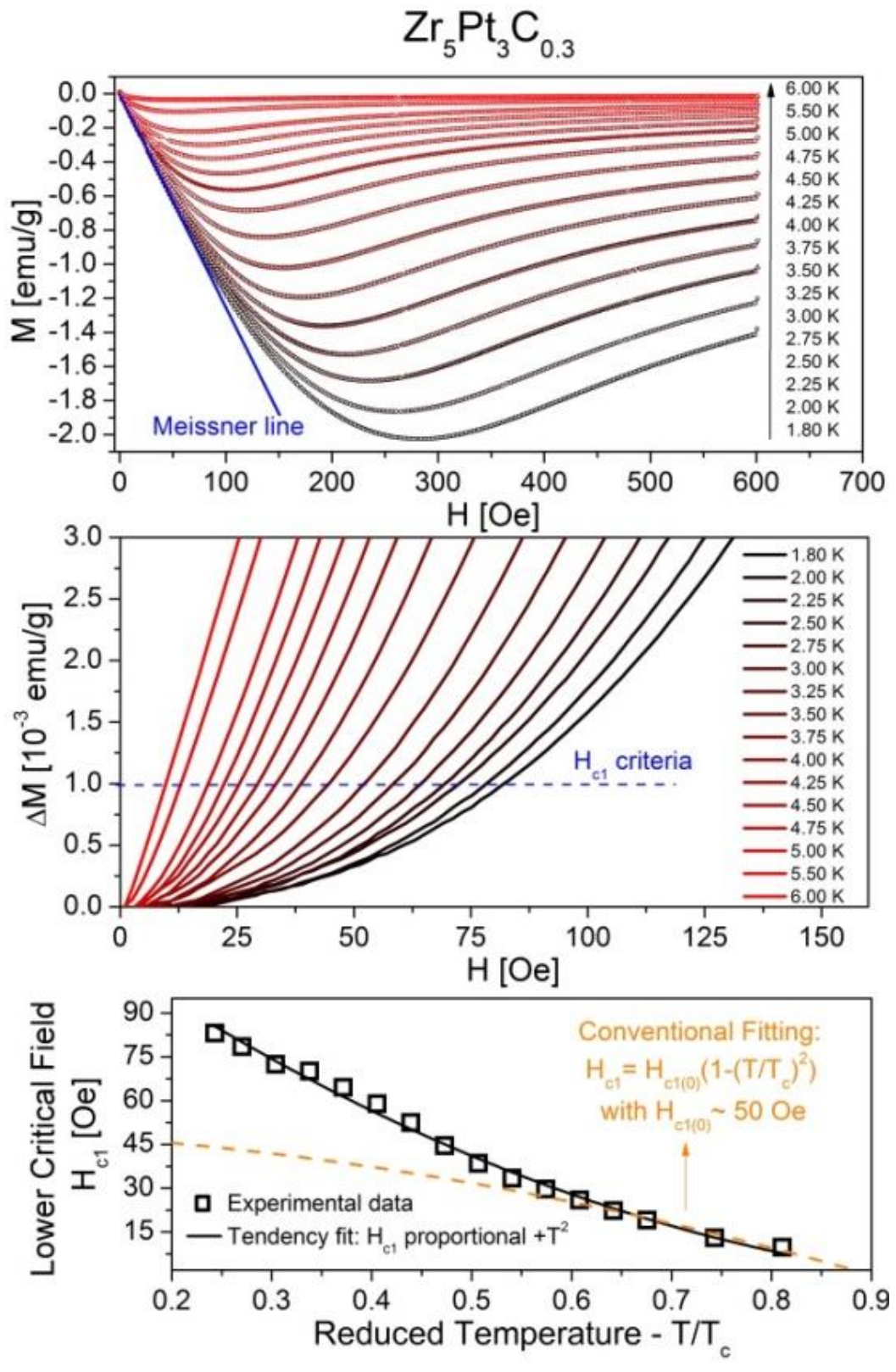

**Fig. 4:** a) Magnetic field dependence of the magnetization at several temperatures, for $Zr_5Pt_3C_{0.3}$. b) Difference to Meissner line and magnetization signal for the criteria of $10^{-3}$ emu/g used to define the lower critical field $H_{c1}$. c) Reduced temperature dependence of lower critical field. Fittings performed using the conventional behavior (orange line) and using a second-order polynomial (black line).

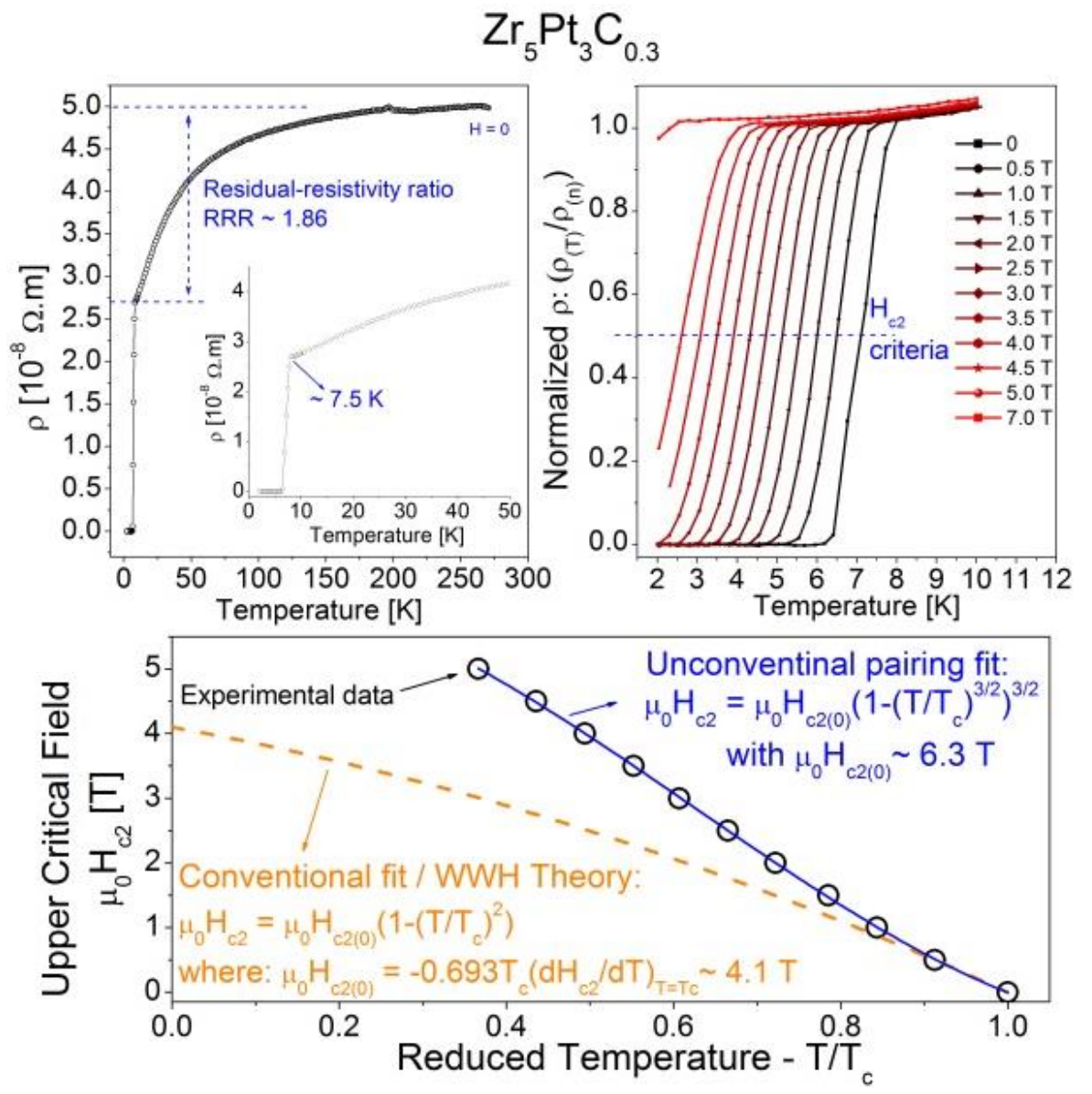

**Fig. 5:** a) Temperature dependence of the resistivity for $Zr_5Pt_3C_{0.3}$. b) Temperature dependence of the normalized magnetoresistivity as a criterion to determine the upper critical field $H_{c2}$. c) Upper critical field versus reduced temperature curves. Fittings performed by isotropic-single band WHH (orange line) and by approaching local pairing mechanism (blue line).

In normal state, the $\rho(T)$ metallic-character was not observed in a range of temperature from $T_c$ to 200 K for $Zr_5Pt_3C_{0.3}$ [Fig. 6a]. Empirical expression for $\rho(T)$ was proposed in terms of three scattering contributions: $\rho_0$ for residual resistivity (impurities/structural defects), $\rho_1 T$ for usual phonon-electron scattering, and $\rho_2 e^{\Delta/T}$ for electron-electron interaction [45], where the exponential is a signature of a characteristic excitation energy for electrons around the Fermi level ($E_F$). The value of $\rho_0$ was estimated as $26.9 \times 10^{-8}$ Ω.m, for higher temperatures the linear-character gives the $\rho_1$ coefficient with the value of $10.6 \times 10^{-11}$ Ω.m, the $\rho_2$ pre-exponential term is $26.5 \times 10^{-8}$ Ω.m, and finally, the $\Delta$ is 29.3 K [Fig. 6b].

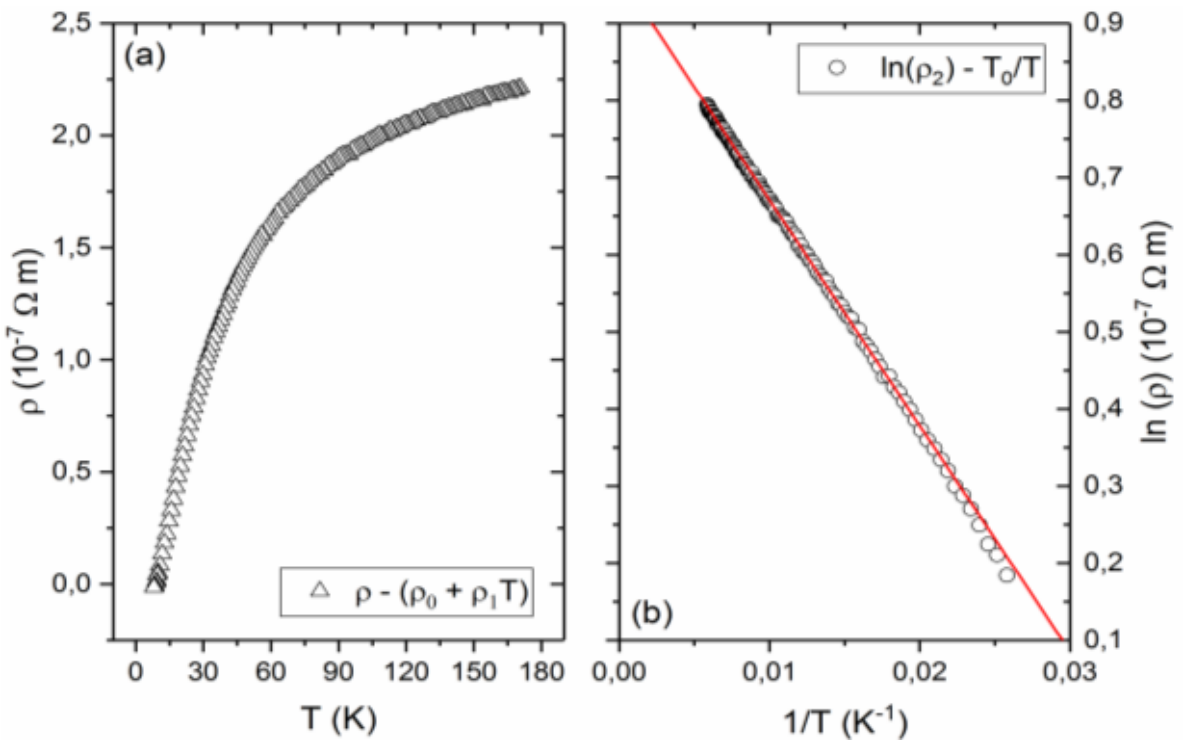

**Fig. 6:** a) Temperature dependence for the exponential term of the resistivity (after subtraction of the linear term and the residual resistivity) for $Zr_5Pt_3C_{0.3}$. b) Inverse-temperature dependence for the logarithm of the exponential term. The fitting (red line) was obtained with the estimate of the pre-exponential coefficient and the $\Delta$.

This behavior reflexes a strong electron-electron interaction on the Fermi surface, in this case, more intense than the usual electron-phonon scattering. The electron localization by electronic interactions is significant in *ρ(T)* behavior, indicating that the electronic structure is neither simply free-electron-like nor completely ionic, but a mixture of both [46]. This phenomenon is a typical signature of compounds with a high density of states (DOS) [47-49].

In order to obtain further insight about the apparent strong-correlation effects due to the anomalous *ρ(T)*, we provide results of the ground-state electronic structure of $Zr_5Pt_3$ from DFT calculations [50,51]. The calculations were performed on the Exciting code using full-potential augmented plane wave with local orbitals [52,53]. Local-density approximation within the prescription introduced by Perdew and Wang was used to treat exchange and correlation effects [54]. A mesh grid with 10 x 10 x 10 *k*–points was used, with muffin-tin radii ($R_{MT}$) and the product $R_{MT}\,K_{Max}$ = 5, where $K_{Max}$ is related with the size of the basis. Experimental values of lattice parameters and atomic positions were used, and spin-orbit coupling was considered.

Both band diagram and Fermi surface showed the presence of 5 bands crossing the $E_F$, leading to a high DOS [Fig. 7a-b]. This may lead to intense electron-electron interactions, in agreement with the presence of the exponential temperature-dependent term of *ρ(T)* in the normal state. In addition, a strong anisotropy in physical properties arises from this three-dimensional complex Fermi surface.

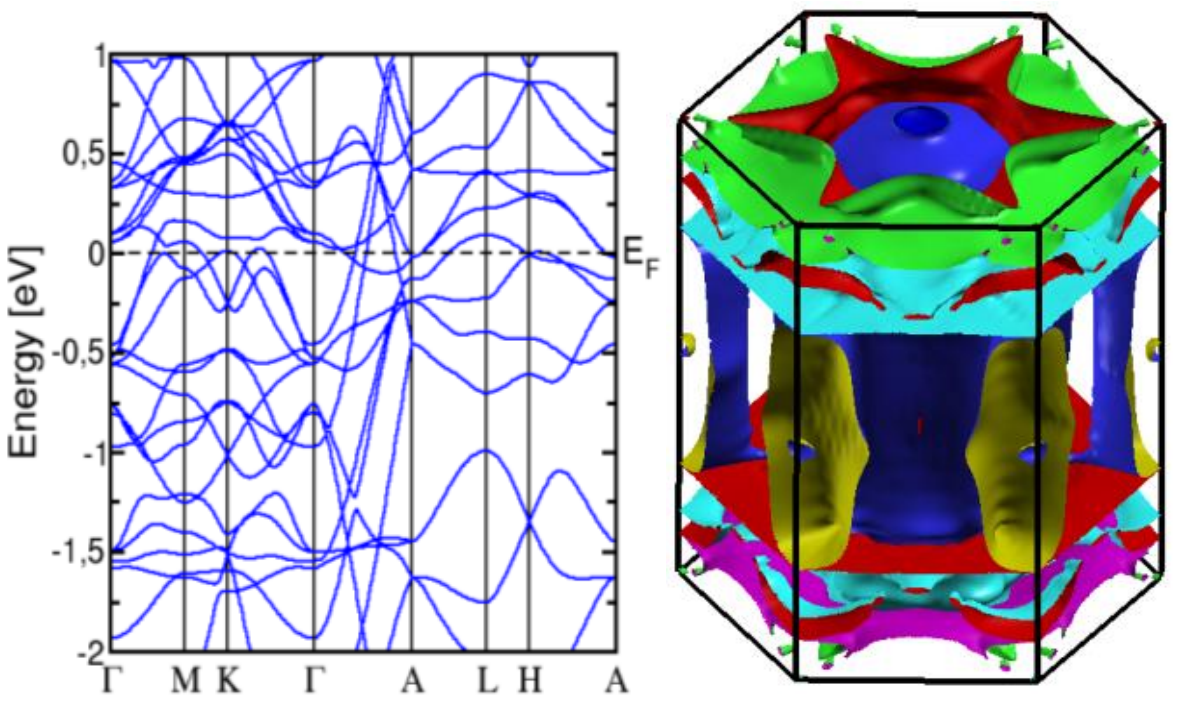

**Fig. 7:** Band structure and Fermi surface of $Zr_5Pt_3$.

For $Zr_5Pt_3$ sample, results show significant contributions at the X-ray photoelectron spectrum (XPS) related to the *d*–*d* hybridizations. Mainly, in the vicinity of $E_F$, when compared with the simulation results for $Zr_5Pt_3$ [Fig. 8]. At the Fermi level, the total $DOS_{cal}$ presents a value of 0.80 states $eV^{-1}.atom^{-1}$ and valence bands below the $E_F$, originated from the contribution of Zr *4d*–states and Pt *5d*–states. DOS contribution can be divided into three parts: From -0.9eV to the conduction band (region I), crossing the Fermi level: the sharp Pt *d*- and *s*-states dominate the shape of DOS. From the bottom, up to the intermediate region of the valence state, it is rather complex, with mixed Zr *d*-state and partly of Pt *p*-hybridized state (region III). The valence bands, with a bandwidth of -4.5 eV, result from a strong hybridization between Pt *d*-state and Zr *d*-state (region II). The wide valley below the $E_F$ implies that the wide separation between bonding and antibonding states is a result of strong Zr–Pt metallic bonds.

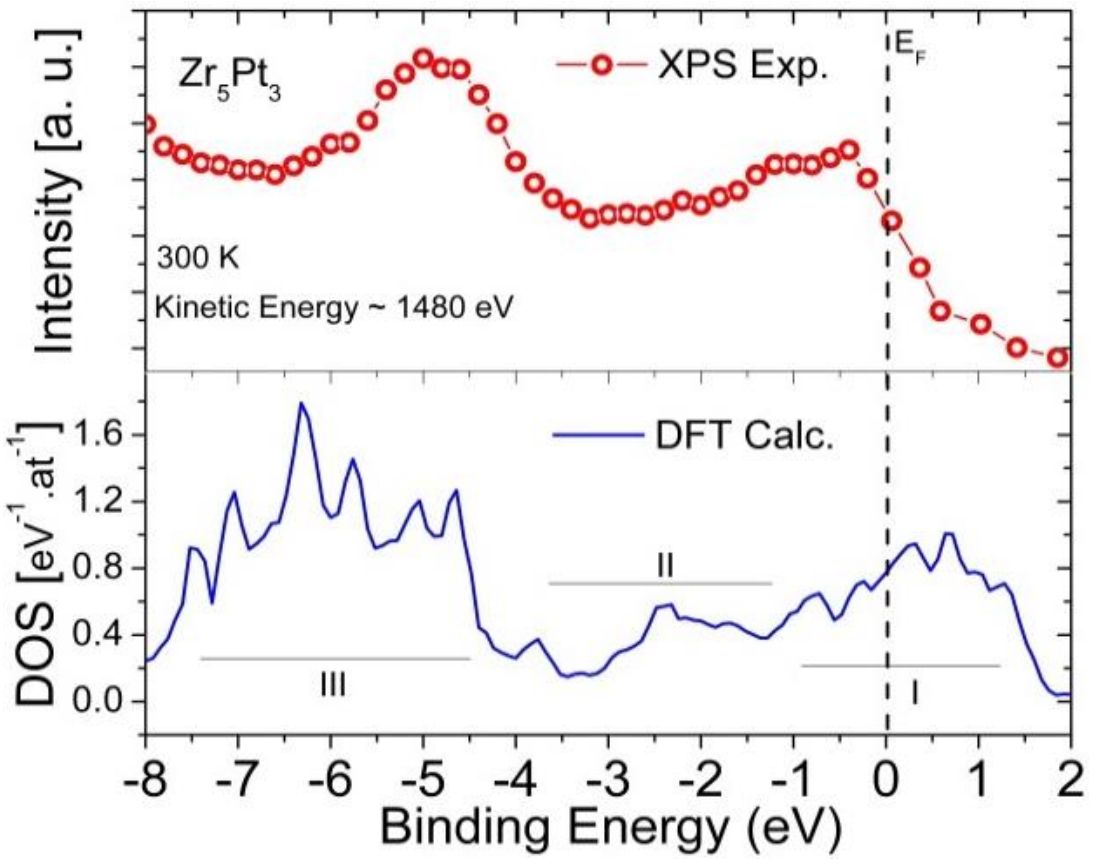

**Fig. 8:** Upper panel) X-ray photoelectron spectrum in the vicinity of the Fermi level for $Zr_5Pt_3$. Lower panel) Density of states obtained by DFT/FP calculations.

Comparison of DOS between $Zr_5Pt_3$ [Fig. 9 upper panel] and $Zr_5Pt_3C_{0.3}$ [Fig. 9 lower panel] was performed by $C_P$ measurements. Consistent with *M(T)* results, $T_c$ appears close to 6.0K and 6.6K as thermodynamic transitions, which agree with bulk superconductivity. The $C_p$ was analyzed with a normal-state fit using the Debye expression: $C_{p(T\to0)} = \gamma T + \sum_{n\geq1}\beta_{2n+1}T^{2n+1}$. The first part reflects the electronic term. Second part is an expansion to phonon contribution. With $\beta_3 = 12/5(N_A K_B \pi^4 \Theta_{D(0)}^{-3})$ for the first-order approximation [see inset Fig. 9a], where $K_B$ is the Boltzmann constant, $N_A$ is the Avogadro number and $\Theta_{D(0)}$ is the initial Debye temperature.

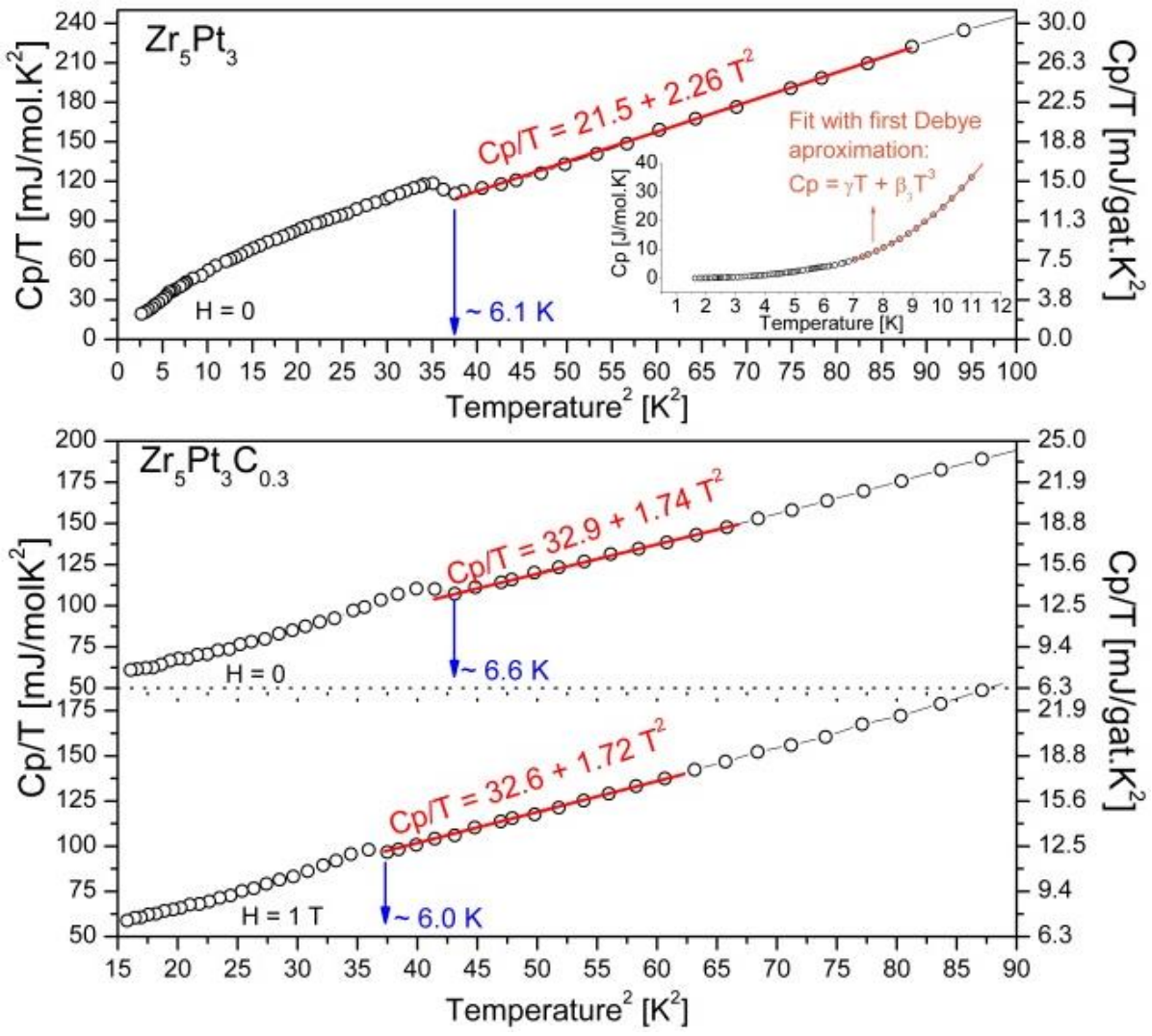

**Fig. 9:** Linearized temperature dependence of the specific heat measurements for $Zr_5Pt_3$ [upper panel] and for $Zr_5Pt_3C_{0.3}$ [lower panel]. The red line is the fit using the Debye theory.

For $Zr_5Pt_3$, we find $\gamma$ = 21.5 mJ/mol.K$^2$ and $\Theta_{D(0)}$ = 190 K. The $\Theta_{D(0)}$ is close to the experimental value of 216 K for $Zr_5Pt_3$ [35], however the $\gamma$-value of 36.8 mJ/mol.K$^2$ (or 4.6 mJ/gat.K$^2$) reported in this reference is questionable (see table 1 of the ref. [35]). Indeed the exam with a linear fit to $C_p$ (see figure 3 of the ref. [35]) indicates another value, close to 19.2 mJ/mol.K$^2$ (or 2.4 mJ/gat.K$^2$), consistent with the value obtained here.

For $Zr_5Pt_3C_{0.3}$ $\gamma \approx 33$ mJ/mol.K$^2$ and $\Theta_{D(0)} \approx 210$ K were obtained, and similar values were found under 1 T. In order to estimate DOS total, we used the equation $\gamma = 1/3(\pi^2 K_B^2(1+\lambda_{EP}))DOS$, where the parabolic dispersion is adjusted by of electron-phonon coupling $(1+\lambda_{EP})$. The average on Fermi surface to $<\lambda_{EP}>$-coupling was obtained by the McMillian expression [55], considering the effective pseudopotential ($\mu$*) equal 0.1 [56]. Despite this expression raise from Eliashberg formulation on s-wave BCS theory, the theoretical arguments are valid too of anisotropic superconductors. Considering $\xi$ value few magnitude orders greater than the interatomic distance and Fermi liquid at the normal state, the determination of $<\lambda_{EP}>$ from experimental $C_p$ remains valid [57]. Consistently with $T_c$ and $\Theta_{D(0)}$ observed for both samples, the $<\lambda_{EP}>$ factor was estimated around 0.73. For $Zr_5Pt_3$, the $<\lambda_{EP}>$ provides the DOS value close to 0.66 states eV$^{-1}$.atom$^{-1}$; coherent with $DOS_{cal}$ (0.80 states eV$^{-1}$.atom$^{-1}$). On the other hand, for $Zr_5Pt_3C_{0.3}$ the DOS increased to around 1.00 states eV$^{-1}$.atom$^{-1}$, indicating an alteration by a factor of 1.5. This clearly reveals that the DOS increase with the C-doping, and it is a probable reason for the pronounced electron-electron interaction on the $\rho(T)$ measurements.

For $Zr_5Pt_3$, the electronic contribution ($C_{p(elec)}$) below $T_c$ [extracted from Fig. 9a] was compared with the BCS exponential-with-$T$ behavior [Fig. 10]. A remarkable divergence from BCS isotropic pairing (red line) was observed above $T_c/T \approx 1.5$. Indeed, the $T$-dependence of $C_{p(elec)}$ shows good agreement with an unusual power law $T^3$ [Fig. 10 inset]. This divergence suggests a nodal symmetry for quasiparticles pairing (QPP). Particularly for $C_{p(elec)}$ the exponential behavior emerge of electronic excitations close to the Fermi level through the $\Delta$-isotropic gap. In the other hand, the existence of nodal QPP carries zero gap regions in Fermi surface with points or lines symmetry. This scenario would explain the unusual $T$ dependence for $Zr_5Pt_3$ comes from a mixture of localized and delocalized QPP, where evidence has shown that nodal gap symmetry can detect for generate quasiparticles from the condensate state [58]. For example, the gap with line nodes conduce to power law dependence $C_{p(ele)} \propto T^2$ and, as observed here to $Zr_5Pt_3$, the $C_{p(ele)} \propto T^3$ for point nodes [7].

In fact, nodal QPP signature was also observed for $Zr_5Pt_3C_{0.3}$ where appears on the linearized temperature-dependence to penetration length ($\lambda_L$) [extracted from $H_{c1}$ diagram - Fig. 4c]. The strong divergence of $\lambda_L(T)$ against to isotropic gap fit (red line) [59] was observed [Fig. 11], and $\lambda_L(T)$ behavior was fitted with an unusual $T^2$-dependence [Fig. 11 inset].

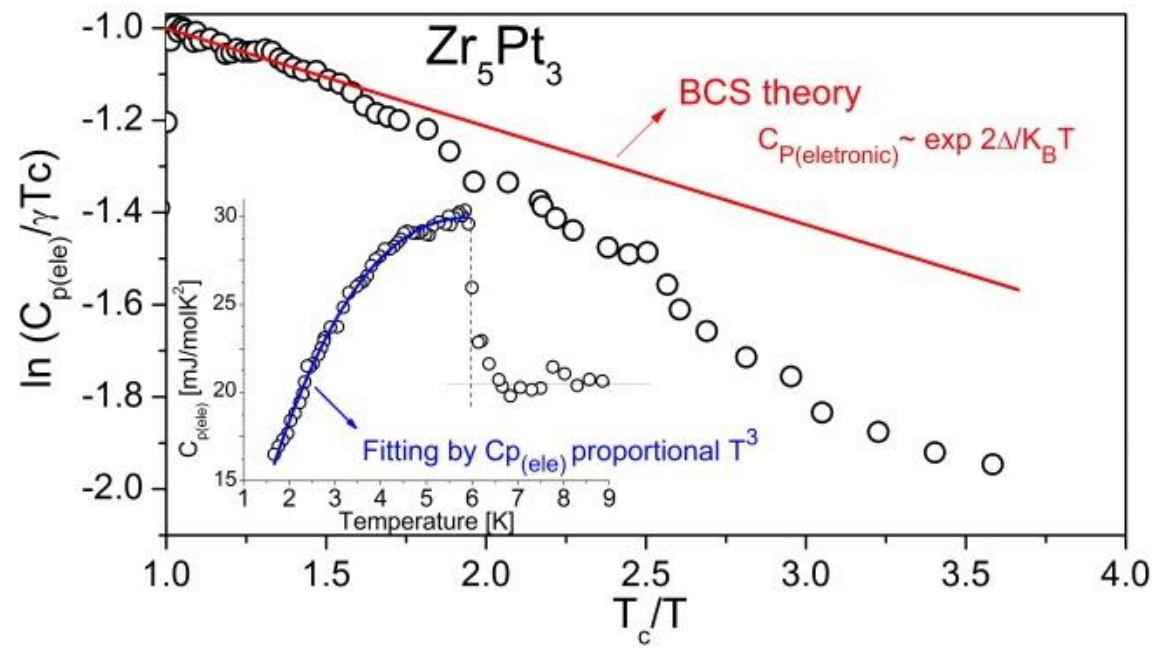

**Fig. 10:** Linearized electronic contribution ln($C_{p(electronic)}/\gamma T_c$) to the $C_p$ [from Fig. 9 upper panel] as a function of $T_c/T$ for $Zr_5Pt_3$. The red line is a fitting based on the BCS theory. The inset presents the fit with $T^3$-dependence power law for $C_{p(electronic)}$ (blue line).

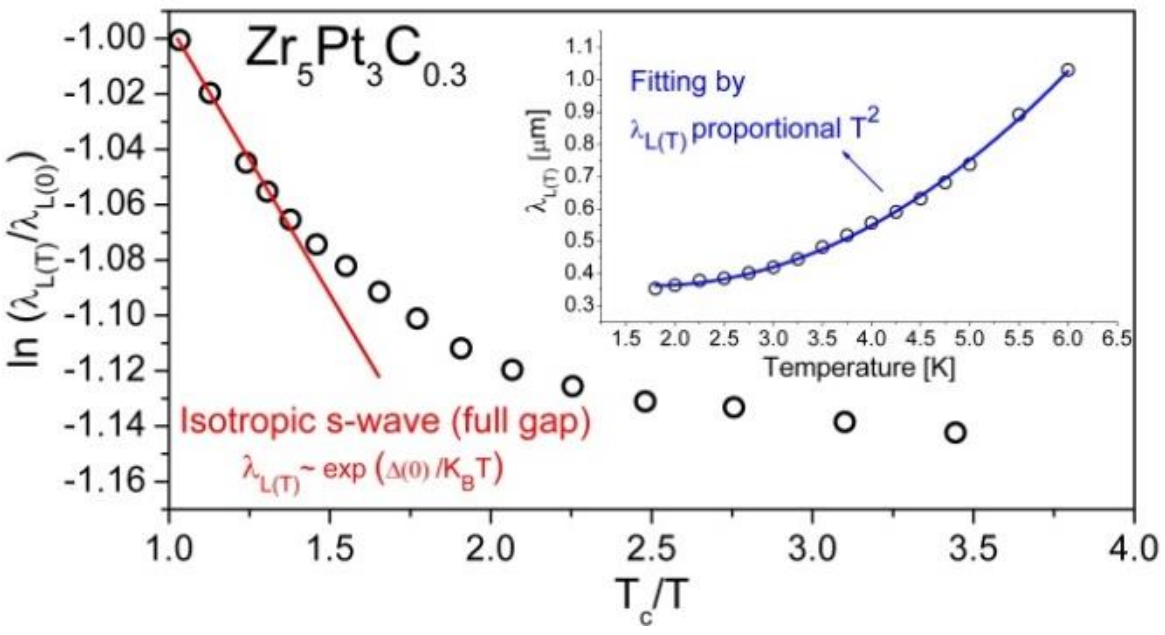

**Fig. 11:** Linearized temperature-dependence of penetration length ln($\lambda_{L(T)}/\lambda_{L(0)}$) [$H_{c1}$ data from Fig. 4c] as a function of $T_c/T$ for $Zr_5Pt_3C_{0.3}$. The red line is a fitting based on the isotropic $s$-wave model [58]. The inset presents the fit with a $T^2$-dependence power-law of $\lambda_{L(T)}$ (blue line).

The $\lambda_L$ was investigated in term of Fermi surface parameters and its anisotropy [4], with the anisotropy on the Fermi velocity ($V_F$) dependence to $\lambda_L$ was presented. In literature the power-law $\lambda_L \propto T^2$ suggest gap asymmetries, goes to existence of points with low energy excitation in QPP nodal scenario [60]. Indeed, on vicinity of mixed state for the d-wave superconductors supercurrents around the vortices nucleus the cause a Doppler shift in the quasiparticle spectrum, which leads to collective excitations of low energy through the nodes [61,62].

Borrowing this reasoning we speculate that with a complex Fermi surface and a high $d$-state contribution of DOS, can generate a anisotropy pairing lead to an unusual power-law dependence at low temperatures. This could be explained by (i) $V_F$ different on anisotropic Fermi surface and (ii) strong coupling in the gapped region (small-$V_F$) and sizable coupling between the small-$V_F$ and large-$V_F$ regions (nodal points). The upward curvature in critical fields will increase as the difference between the Fermi velocities increases. Thus, together with the anomalous behavior of $H_{c1}$ and $H_{c2}$, this surprising result can suggest the existence of an unconventional superconductivity scenario [63-66].

In conclusion, a previously unappreciated set of physical properties is intrinsic to the $Zr_5Pt_3C_y$. Unprecedented results showed $D8_8$ compounds as a promising materials group with very interesting properties. Initial results showed unambiguously that $Zr_5Pt_3C_{0.3}$ is a bulk superconductor. Several experimental evidences reveal a strong divergence from conventional behavior of BCS theory. Normal-state $\rho$ and $C_p$, together with DFT calculations, suggest the presence of intense electron-electron interactions around the Fermi level due to high DOS. The signatures of strong electronic correlations and nodal superconductivity, presents an intriguing scenario for condensed-matter.

This material is based upon work is supported by MINECO (MAT2016/75955) and FAPESP (14/25235-3; 16/10167-8; 16/11565-7). Work at the National Nanotechnology Laboratory at Campinas (Proj. 12863 and 13555).